# Nanosecond Optically Induced Phase Transformation in Compressively Strained BiFeO$_3$ on LaAlO$_3$

Youngjun Ahn,[1] Anastasios Pateras,[1] Samuel D. Marks,[1] Han Xu,[2] Tao Zhou,[3] Zhenlin Luo,[2] Zuhuang Chen,[4] Lang Chen,[5] Xiaoyi Zhang,[6] Anthony D. DiChiara,[6] Haidan Wen,[6] and Paul G. Evans[1,†]

[1] *Department of Materials Science and Engineering, University of Wisconsin-Madison, Madison, Wisconsin 53706, USA*

[2] *National Synchrotron Radiation Laboratory, University of Science and Technology of China, Hefei, Anhui 230026, China*

[3] *ID01/ESRF, 71 Avenue des Martyrs, 38000 Grenoble Cedex, France*

[4] *School of Materials Science and Engineering, Harbin Institute of Technology, Shenzhen 518055, China*

[5] *Department of Physics, Southern University of Science and Technology, Shenzhen, Guangdong 518055, China*

[6] *Advanced Photon Source, Argonne National Laboratory, Argonne, Illinois 60439, USA*

[†]pgevans@wisc.edu

**Abstract**

Above-bandgap optical illumination of compressively strained BiFeO$_3$ induces a transient reversible transformation from a state of coexisting tilted tetragonal-like and rhombohedral-like phases to an untilted tetragonal-like phase. Time-resolved synchrotron x-ray diffraction reveals that the transformation is induced by an ultrafast optically induced lattice expansion that shifts the relative free energies of the tetragonal-like and rhombohedral-like phases. The transformation proceeds at boundaries between regions of the tetragonal-like phase and regions a mixture of tilted phases, consistent with the motion of a phase boundary. The optically induced transformation demonstrates that there are new optically driven routes towards nanosecond-scale control of phase transformations in ferroelectrics and multiferroics.

Ferroelectric and multiferroic oxides can be transformed between structural phases with different structures and properties by applied pressure, stress, and electric or magnetic fields [1-3]. These systems are particularly sensitive to external stimuli when placed near a phase boundary by lattice-mismatched epitaxial growth or chemical substitution [4-6]. Epitaxial growth on an $LaAlO_3$ (LAO) substrate, for example, places multiferroic $BiFeO_3$ (BFO) near the boundary between rhombohedral-like and tetragonal-like phases that differ significantly in their properties [4]. Both phases have monoclinic symmetry but have different directions and magnitudes of the ferroelectric polarization [7,8]. Furthermore, the magnetism of the tetragonal-like phase exhibits weaker order than the rhombohedral-like phase due to the suppression of the canting of the antiferromagnetic sublattice [9,10]. Differences in the Fe-O bond lengths and Fe-Fe distances also lead to differences in the band structure and in electronic and optical properties of the two phases [11]. Understanding the mechanisms of the transformation and how it can be induced at ultrafast timescales has the potential to lead to the creation of optically tunable and reconfigurable complex oxide electronic and optical materials.

A reversible transformation to the tetragonal-like phase of BFO can be driven by an external electric field [12,13]. Density functional theory studies indicate that the tetragonal-like phase becomes energetically favorable when the lattice is expanded along the out-of-plane direction [14]. Several questions remain regarding the transformation mechanisms, the transformation pathway, and its dynamics. A challenge in studying the transformation using an electric fields is that the time resolution is limited by the charging times of the relatively large thin-film devices and with difficulty in applying high fields in electrically leaky materials [14].

Femtosecond-duration above-bandgap optical pulses generate a transient out-of-plane strain in ferroelectrics and multiferroics, providing a non-contact, ultrafast means to investigate their dynamics. Optical excitation of ferroelectrics and multiferroics generates lattice expansion up to on the order of 1% [15-19]. One mechanism of the expansion arises from screening of surface and interfacial bound charges by excited charge carriers, changing the internal electric field and generating electromechanical distortion [19,20]. Studies of the electrically driven BFO phase transformation indicate that the optically excited expansion would be sufficiently large to induce the transformation to the tetragonal-like phase [14].

Here we report a rapid reversible transformation between phases in compressively strained BFO induced by ultrafast optical excitation. Time-resolved x-ray microdiffraction shows that the optically induced lattice expansion and the transformation both reach their maxima within 1 ns after excitation. X-ray microscopy shows that the transformation occurs in regions in which there are boundaries between the tetragonal-like phase and the tilted tetragonal-like and tilted rhombohedral-like phases. Together, these observations suggest a mechanism in which the tetragonal-like phase is stabilized by the lattice expansion and grows into regions of coexisting populations of the tilted tetragonal-like and tilted rhombohedral-like phases. The structural changes during the optically induced transition are distinct from those at high temperature, including differences in the variation of the lattice parameter and phase population.

Time-resolved x-ray microdiffraction experiments were conducted at station 7ID-C of the Advanced Photon Source using the arrangement in Fig. 1(a). X-rays with a photon energy of 11 keV and 100 ps pulse duration were focused to a 400 nm full-width-at-half-maximum (FWHM) probe spot. The optical pump consisted of 50-fs-duration pulses with 1 kHz repetition rate and 400 nm wavelength, above the bandgap of BFO on LAO [21-23]. A complementary full-field x-ray diffraction microscopy study employed beamline ID-01 of the European Synchrotron Radiation Facility [24].

The 70 nm-thick BFO thin film was grown on an $La_{0.7}Sr_{0.3}MnO_3$ (LSMO) bottom electrode on LAO by pulsed laser deposition [12]. The tetragonal-like and rhombohedral-like phases are referred to as the T and R phases, respectively [4]. The epitaxial mismatch of -4.5% between BFO and LAO leads to the formation of the T phase, the stable form at high compressive strain. The stored elastic energy of the film is reduced by forming regions of stripes of alternating tilted-T (TT) and tilted-R (TR) phases with widths of approximately 50 nm within the larger field of the T phase [4,8,25-27]. Figures 1(b)-(d) show the diffracted intensity near the 002 T phase reflection acquired with an unfocused 50 μm FWHM x-ray beam, averaging over all BFO phases. Figure 1(b) shows part of the $Q_z$-$Q_y$ plane of reciprocal space in which the TT and TR phases appear at $Q_z = 2.69$ Å$^{-1}$ and 3.02 Å$^{-1}$, respectively. Eight intensity maxima arising from the TT phase and a single T reflection appear in the $Q_y$-$Q_x$ plane at $Q_z = 2.70$ Å$^{-1}$ in Fig. 1(c). The TR phase also exhibits four pairs of reflections, as in Fig. 1(d). The eight intensity maxima of the TT and TR phases indicate that there are eight domains of each tilted phase [12,28,29]. The reflections in Figs. 1(c) and (d) are numbered in order to discuss them precisely. In the striped microstructure, the TT reflections (1) and (4) are paired with the TR reflections (5) and (8), respectively [27,30,31].

Optical excitation leads to an expansion of the T phase, as in Fig. 1(e), in which the 002 T phase reflection measured using x-ray microdiffraction is shifted to a lower $Q_z$ at 1 ns after excitation at an fluence of 10.7 mJ/cm$^2$. At this fluence, the lattice expansion has a magnitude of 0.16% and is accompanied by an 8% increase in the integrated diffracted intensity. The changes in the volume of the film occupied by each phase were measured using the changes in the integrated intensities of corresponding x-ray reflections. The phase populations were determined using an analysis considering the different scattering factors of each phase (see Supplemental Material [32]). The T, TT, and TR phases initially occupy 75%, 15%, and 10% of the film, respectively, matching report for BFO films on LAO with similar thicknesses [4]. The intensities of the TT and TR reflections decreased by 28% and 7%, respectively, at 1 ns with a fluence of 10.7 mJ/cm$^2$. The population changes indicate that there is a transformation from the TT and TR phases to the T phase.

The variation of the out-of-plane lattice parameter of the T phase as a function of the delay time $t$ is shown in Fig. 2(a) for a fluence of 10.7 mJ/cm$^2$. The out-of-plane lattice expansion of the T phase was 0.14% at $t$=200 ps and reached a maximum of 0.16% at $t$=1 ns. Two physical effects are apparent in the expansion of the T phase. The initial expansion occurs with a timescale set by the propagation of a longitudinal elastic wave through the film thickness, on the order of tens of ps, which is less than the experimental time resolution [17]. A second, long-timescale, component of the lattice expansion, apparent from 200 ps to 1 ns in Fig. 2(a), is not consistent with the acoustic response and is compared below with the timescale of the motion of the phase boundaries. The possibility that thermal diffusion produces the slow risetime can be evaluated by comparing the thermal diffusion length $\sqrt{Dt}$ with the film thickness. Using $D$=1.2×10$^{-4}$ m$^2$/s [19] gives $\sqrt{Dt}$=350 nm at $t$=1 ns, which is far larger than the BFO thickness and indicates that the peak average temperature is reached at a far shorter time than 1 ns. A complete thermal diffusion simulation based on methods in ref. [39] appears in the Supplementary Material.

The integrated intensity of the T phase reflection, Fig. 2(b), increases for times up to 1 ns. The intensities of the TT (4) and TR (8) reflections decrease after optical excitation and reach minima at time $t$=1 ns, matching the timescale of the change in the T phase intensity. At $t$=1 ns, the population of the T phase increases to 81% of the BFO volume, and gradually returns to 76% at $t$=12 ns. The populations of the TT and TR phases decrease to 10% and 9% at $t$=1 ns, respectively, in agreement with the 6% change in the T phase population at that time. In contrast with this mixed-phase sample, optical excitation of pure R phase BFO leads to negligible intensity changes [17-19].

The magnitude of the T phase lattice expansion increases at higher fluence, reaching 0.45% at $t=1$ ns at 20 mJ/cm$^2$, as in Fig. 3(a). The intensities of the T, TT, and TR phases also exhibit larger-magnitude changes as a function of increasing fluence, as in Fig. 3(b). The monotonic variation of the intensity changes is consistent with a continuous shift of the energetic balance between the competing phases rather than a transformation at a critical value of the optically induced expansion. A similar continuous transformation with increasing electric field is observed in the electric-field-driven transformation from R to T phases of BFO [14].

A Landau-Ginzburg-Devonshire model was used to predict the energetic stability of the T and R phases as a function of the optically induced distortion. The phases have equal free energies at a biaxial misfit strain of -4.3%, consistent with the previously reported phase boundary. An out-of-plane lattice increases the free energy of the T phase less than the R phase (see Supplementary Material) and favors a transformation from the R phase to the T phase. The difference in the free energies grows monotonically with lattice expansion. The continuous increase in the free-energy difference agrees with the experimental observation that the transformation proceeds without a single threshold value of the fluence or expansion.

The differences between optical excitation and heating were evaluated using variable-temperature diffraction with a laboratory x-ray source. As shown in Fig. 3(c), the T phase lattice parameter expands by 0.45% from room temperature to 110 ºC and shrinks as the temperature increases from 110 ºC to 200 ºC, where it has a value close to the lattice parameter at 90 ºC. The variation of the intensities during heating, Fig. 3(d), does not match the intensity changes resulting from the optically induced transformation. First, heating leads to a decrease (rather than the optically induced increase) in the T phase intensity, by 20% at 110 ºC. A similar intensity decrease has been observed in pure T phase BFO, which indicates that the decrease in the intensity of the T phase reflection at elevated temperatures does not originate from a phase transformation [40]. Similarly, the intensities of the TT and TR phases are not consistent with a transformation between phases below 130 ºC. The change in integrated intensity of the TR phase reflection is negligible during heating to 130 ºC. The TT phase reflection intensity increases by 10% at 70 ºC and drops to its room-temperature value at 130 ºC.

The transformation observed at high temperature is also distinct from the optically induced transformation. From 140 ºC to 200 ºC, the intensities of the TT and TR phase reflections decrease and the intensity of the T phase reflection increases, matching literature reports [25,26]. A key difference between optical and thermal effects, however, is that the high-temperature phase transformation occurs in a regime in which the lattice parameter of the T phase decreases, opposite to the optically excited observation. Similarly, the initial increase in the T phase lattice parameter during heating is accompanied by a decrease in the T phase intensity, rather than the increase observed during optically induced expansion.

The spatial arrangement of the coexisting BFO phases suggests a mechanism for the transformation. Scanning x-ray microdiffraction images reveal that the TT and TR phases exist in multiple spatially separated populations and that the optically induced transformation occurs via the simultaneous changes in the coexisting variants. Microdiffraction maps of the integrated intensities of TT (4) and TR (8) phase reflections, Fig. 4(a), show that these variants occur in the same region, matching previous reports [29]. A similar spatial correlation is also exhibited by the TT (1) and TR (5) variants and is observed in full-field x-ray microscopy (see Supplementary Material).

A time-resolved microscopy study was conducted to probe the spatial variation of the optically induced transformation. The lattice expansion in the T phase was measured using reciprocal space maps acquired in a 1.2 µm × 200 µm area at the position indicated by arrows in Fig. 4(a) before optical excitation and at $t=1$ ns following pulses with a fluence of 10.7 mJ/cm$^2$. The optically induced T-phase lattice expansion in Fig. 4(b) is independent of position, which indicates that the expansion does not depend on the local phase population.

The optically induced change in intensity and the extent of the optically induced transformation depend very strongly on the local phase population. The T phase intensity and the fractional change in intensity after optical excitation both vary significantly as a function of position, as shown in Figs. 4(c) and 4(d). In contrast with the nearly constant optically induced expansion, the fractional change in T phase intensity ranges from 0 to nearly 20%. Regions with near-zero change in the T-phase intensity exhibit two key structural features: (i) these regions have low TT and TR phase population (see Supplemental Material) and (ii) the T phase lattice parameter, Fig. 4(b), is large and has a value close to the lattice parameter of films in which the T phase has not elastically relaxed through the formation of other phases [41]. Taken together, these observations indicate that the transformation proceeds only in regions where T, TT, and TR phases coexist.

The observation here of a simultaneous decrease in TT and TR phase reflections following optical excitation is not consistent with a previously proposed sequential transformation mechanism involving an initial transition between the TR and TT phases, followed by a subsequent transformation to the T phase [31]. The simultaneous and proportional changes in the intensities of x-ray reflections of spatially coexisting tilted phases instead suggest a mechanism in which the tilted phases are simultaneously transformed to the T phase. A mesoscopic process that would enable the simultaneous and continuous transformation from a mixture of TT and TR phases to the T phase is illustrated in Fig. 4(e). In this process, the boundary between the T phase and the TT and TR phases progresses into the mixed phase region. A similar growth of the T phase into mixed-phase regions has been observed using band-excitation piezoelectric force microscopy, with nearly equal magnitudes of the transient T-phase lattice expansion and a transformation over lateral distances consistent with the changes in TT and TR phase populations reported here [42].

The observation of the optically induced transformation in BFO has several implications in understanding and applying phase transformations in ferroelectrics and related materials. With respect to electromechanical distortion, the results suggest that nanosecond-scale lattice expansion with a large contribution arising from the transformation between phases can be produced by optical excitation. Furthermore, the optically induced phase transformation can potentially suppress the magnetic moments of BFO because the transformed part of the film changes from the R phase, which has a large magnetic ordering, to pure T phase with weak magnetic order [10]. This work ultimately broadens the potential of ultrafast optical excitation to change structural phases and coupled multiferroic properties.

This work was supported by the US National Science Foundation through grant number DMR-1609545. H. W. acknowledges the support of U.S. Department of Energy, Office of Science, Basic Energy Sciences, Materials Sciences and Engineering Division, for instrumentation development of time-resolved x-ray microdiffraction. This research used resources of the Advanced Photon Source, a U.S. Department of Energy Office of Science User Facility operated for the DOE Office of Science by Argonne National Laboratory under Contract No. DE-AC02-06CH11357. Z. H. C. acknowledges the National Science Foundation of China (No. 51802057) and a startup grant from Harbin Institute of Technology, Shenzhen, China, under project number DD45001017. Z. L. Luo and H. Xu were supported by the National Key Basic Research Program of China (2016YFA0300102) and the National Natural Science Foundation of China (11434009, 11675179, U1532142). L. C. acknowledges the Science and Technology Research Items of Shenzhen (JCYJ20170412153325679 and JCYJ20180504165650580).

FIG 1. (a) X-ray diffraction experimental schematic illustrating the directions of the optical and x-ray pulses and the mesoscale arrangement of coexistenting TT, TR, and T phases. Arrows illustrate the growth of the T phase region by phase boundary motion. (b) Section of the $Q_y$-$Q_z$ plane of reciprocal space at $Q_x = 0$, near the 002 T phase reflection. (c) TT and T phase reflections in the $Q_x$-$Q_y$ plane at $Q_z = 2.70$ Å$^{-1}$. (d) TR phase reflections in the $Q_x$-$Q_y$ plane, integrating from $Q_z = 2.90$ Å$^{-1}$ to 3.08 Å$^{-1}$. Individual reflections are numbered. (e) Intensity as a function of $Q_z$ for the 002 T phase reflection before optical excitation and at $t=1$ ns with fluence 10.7 mJ/cm$^2$.

FIG 2. (a) Lattice expansion in the T phase and (b) fractional intensity changes of T, TR (8), and TT (4) phase reflections as a function of time following optical excitation at fluence 10.7 mJ/cm$^2$.

FIG 3. (a) T phase lattice expansion and (b) integrated intensities of the T, TT (1), TT (4), TR (5) and TR (8) reflections as a function of fluence at $t=1$ ns. Angular scans recording the intensities were conducted for different numbers of values of the fluence for each phase, resulting in different numbers of fluence points for each reflection. (c) T phase lattice expansion and (d) integrated intensities of the T, TT, and TR phase reflections as a function of temperature.

FIG 4. (a) Scanning x-ray microdiffraction maps of TT (4) and TR (8) phases. (b) Lattice expansion along the out-of-plane direction, out-of-plane lattice parameters, (c) integrated intensities before optical excitation and at $t=1$ ns, and (d) fractional T phase intensity change along the line indicated by black arrows in (a). Shaded areas in (c) and (d) indicate regions with large T phase lattice parameter and small changes in the T phase population. (e) Top-view schematic of an optically induced phase transformation mechanism involving expansion of the T phase into the region of mixed TT and TR phases.

Ahn *et al*., Figure 1.

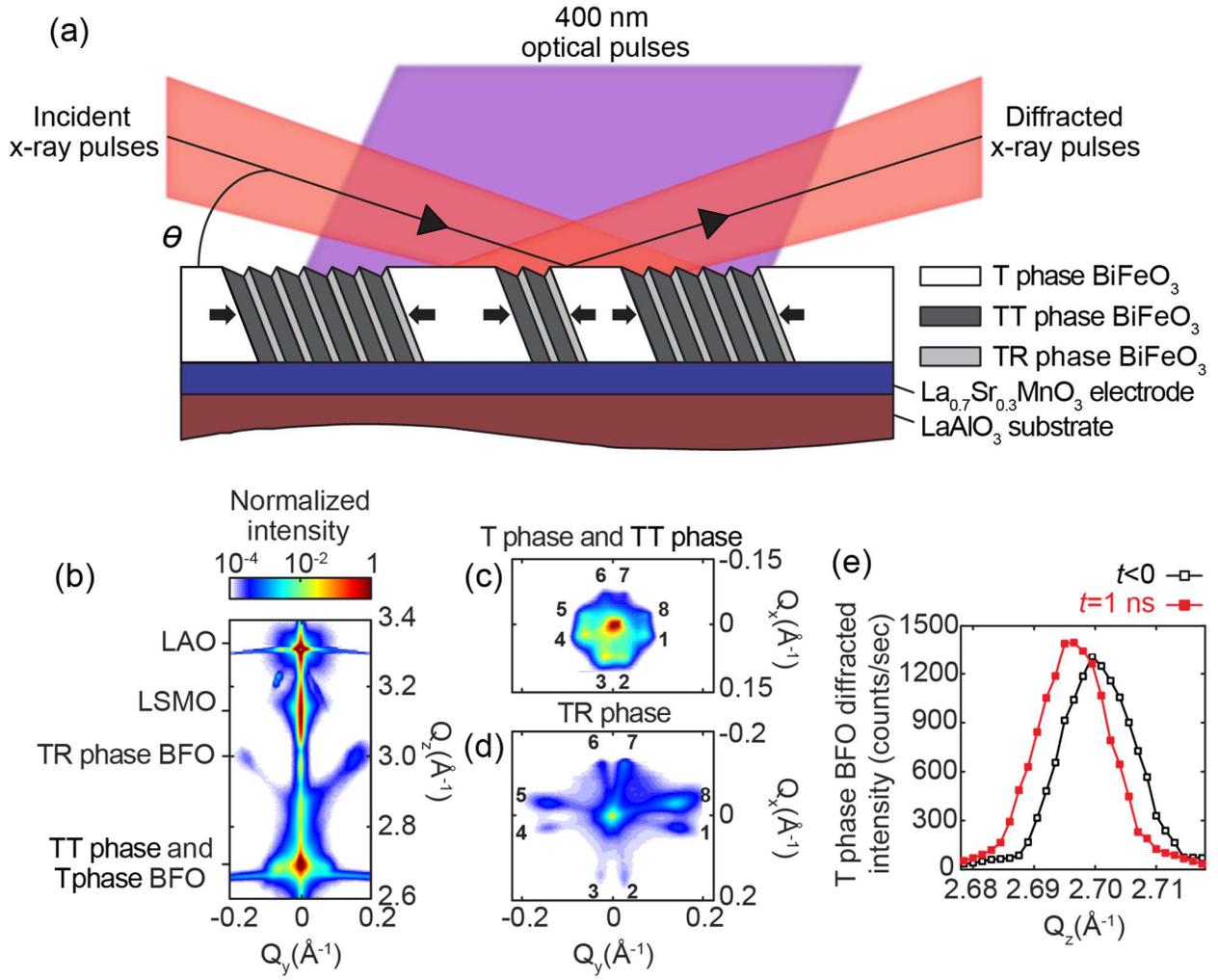

Ahn *et al*., Figure 2.

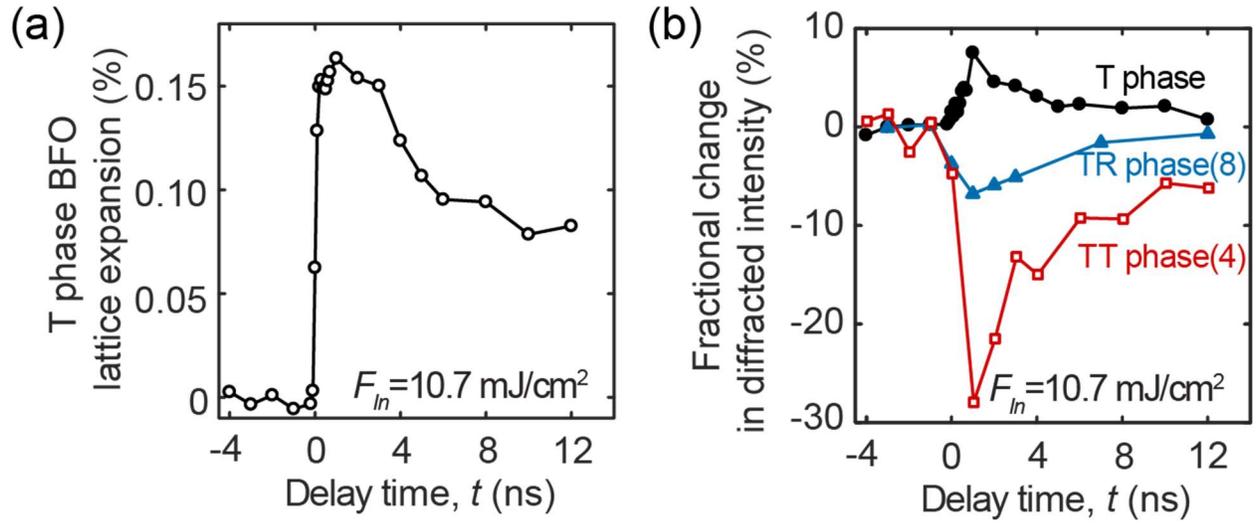

Ahn *et al*., Figure 3.

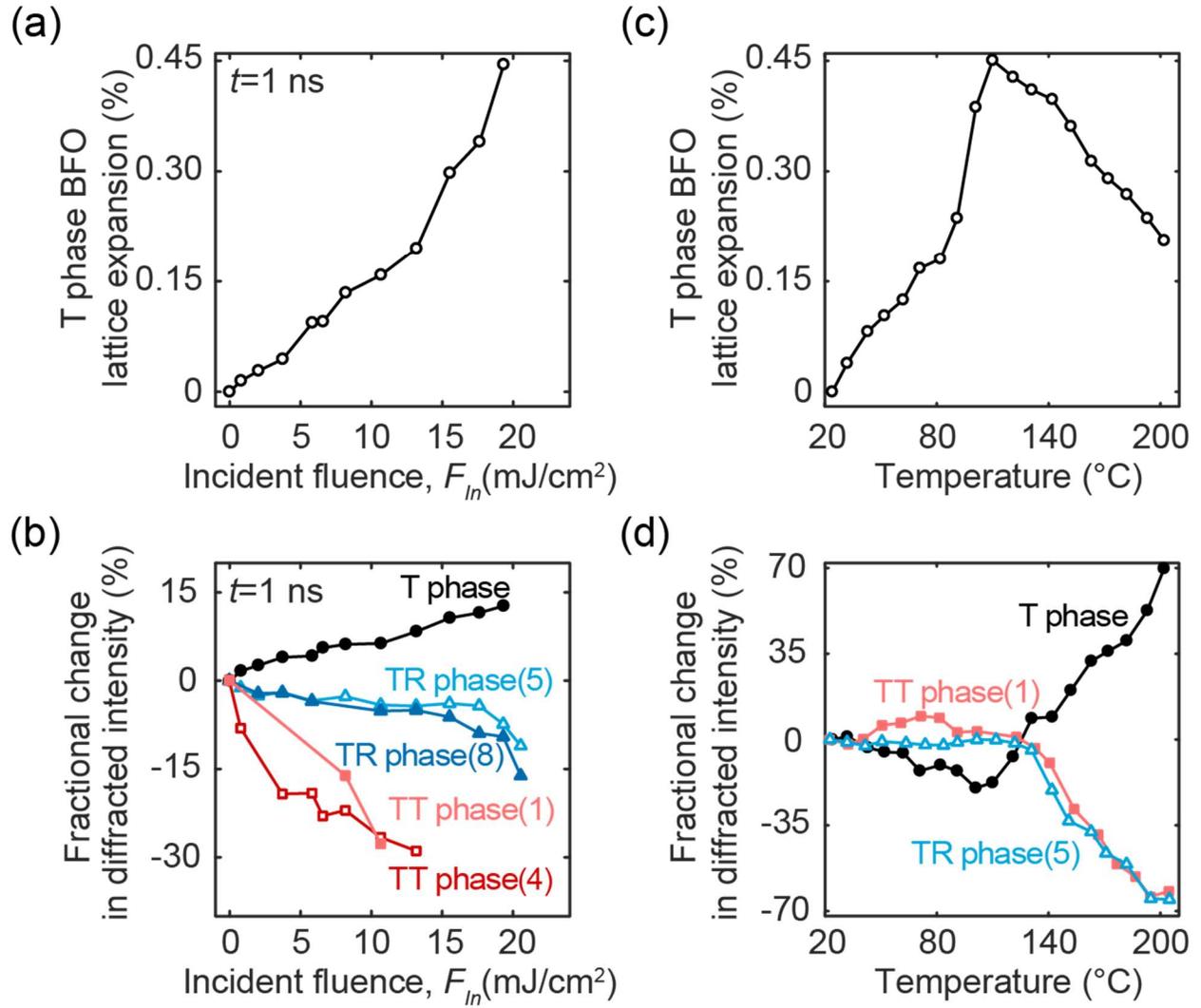

Ahn *et al*., Figure 4.

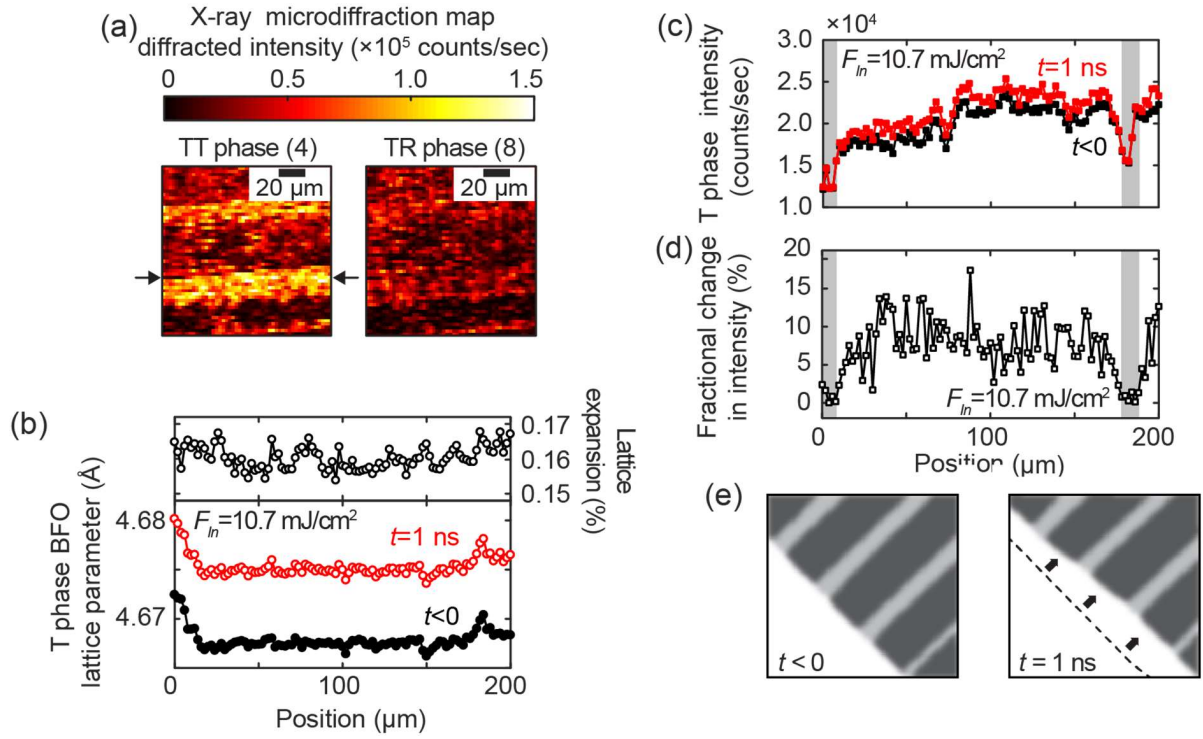

# Supplemental Material for "Nanosecond Optically induced Phase Transformation in Compressively Strained BiFeO$_3$ on LaAlO$_3$"

Youngjun Ahn,[1] Anastasios Pateras,[1] Samuel D. Marks,[1] Han Xu,[2] Tao Zhou,[3] Zhenlin Luo,[2] Zuhuang Chen,[4] Lang Chen,[5] Xiaoyi Zhang,[6] Anthony DiChiara,[6] Haidan Wen,[6] and Paul G. Evans[1,†]

[1]Department of Materials Science and Engineering, University of Wisconsin-Madison, Madison, Wisconsin 53706, USA

[2]National Synchrotron Radiation Laboratory, University of Science and Technology of China, Hefei, Anhui 230026, China

[3]ID01/ESRF, 71 Avenue des Martyrs, 38000 Grenoble Cedex, France

[4]School of Materials Science and Engineering, Harbin Institute of Technology, Shenzhen 518055, People's Republic of China

[5]South University of Science and Technology of China, Shenzhen 518055, People's Republic of China

[6]Advanced Photon Source, Argonne National Laboratory, Argonne, Illinois 60439, USA

[†]pgevans@wisc.edu

## 1. Experimental details

X-ray microdiffraction experiments at the Advanced Photon Source employed Fresnel zone plate x-ray focusing optics. Diffracted x-ray photons were detected using a pixel-array detector (Pilatus 100K, Dectris Ltd.).

Full-field imaging experiments at the European Synchrotron Radiation Facility used a photon energy of 8 keV and a 50-element Be compound refractive lens objective. Images were recorded using a charge-coupled device (Zyla 5.5 sCMOS, Andor, Inc.) placed at 3 m from the sample, yielding a 250 nm effective pixel size. The resolution along the beam direction was increased to 650 nm due to the x-ray projection angle.

## 2. Phase distribution of T phase and TT and TR phases in multi-phase BFO on LAO

The population of structural phases in thin BFO film on LAO was measured by constructing x-ray reciprocal space maps using the microfocused synchrotron x-ray beam. The map was constructed from x-ray diffraction patterns obtained at x-ray incident angles ranging from 13º to 17º. Figure S1(a) shows the region of the $Q_y$-$Q_z$ plane of the reciprocal space map at $Q_x = 0$ Å$^{-1}$ near the 002 T- phase reflection. One TT and one TR phase reflections are apparent in Fig. S1(a). Note that the microdiffraction map in Fig. S1 covers a smaller area of the sample than the maps collected with an unfocused 50 μm beam shown in Figs. 1(b), (c), and (d) of the text.

As shown in Figs. S1(b) and (c), two TT phase and two TR phase reflections arise at reciprocal-space locations indicated by red dotted boxes. The three-dimensional diffraction patterns were integrated to measure the population of the TT phase and TR phases within the probed area. The range of the integration spanned a range in $Q_z$ from 2.65 Å$^{-1}$ to 2.78 Å$^{-1}$ for the TT phase and from 2.90 Å$^{-1}$ to 3.08 Å$^{-1}$ for the TR phase.

The intensity distributions in the $Q_x$-$Q_y$ planes were integrated along $Q_x$ or $Q_y$ so that the line profiles of the intensity of four tilted phase reflections are obtained, as shown in Fig. S1(d). The intensities from the tails of the T phase reflection and truncation rod were then subtracted. The T phase reflection provides 74% of total diffracted intensity of the region probed by microfocused x-ray beam. The TT phase reflections and TR phase reflections provide 14% and 12% of the total diffracted intensity.

The structure factor of T phase at 11 keV of x-ray energy is smaller than the structure factor of the R phase by 8.0%, based on the atomic positions given in Refs. [1-3]. Therefore, the diffracted intensity of T-like phase reflection is a factor of 1.17 weaker than that of R phase reflection when the number of unit cells of T phase and R phase are the same. In this calculation, the structure of R phase is assumed to be the same as the structure of rhombohedral BFO. In addition, we assume that the structure factors of T phase and TT phase are the same. With these assumptions, the initial populations of T, TT and TR phases within the probed region are 75%, 15%, and 10%, respectively. The total area fraction of 25% of the TT and TR phases matches a previous report [4].

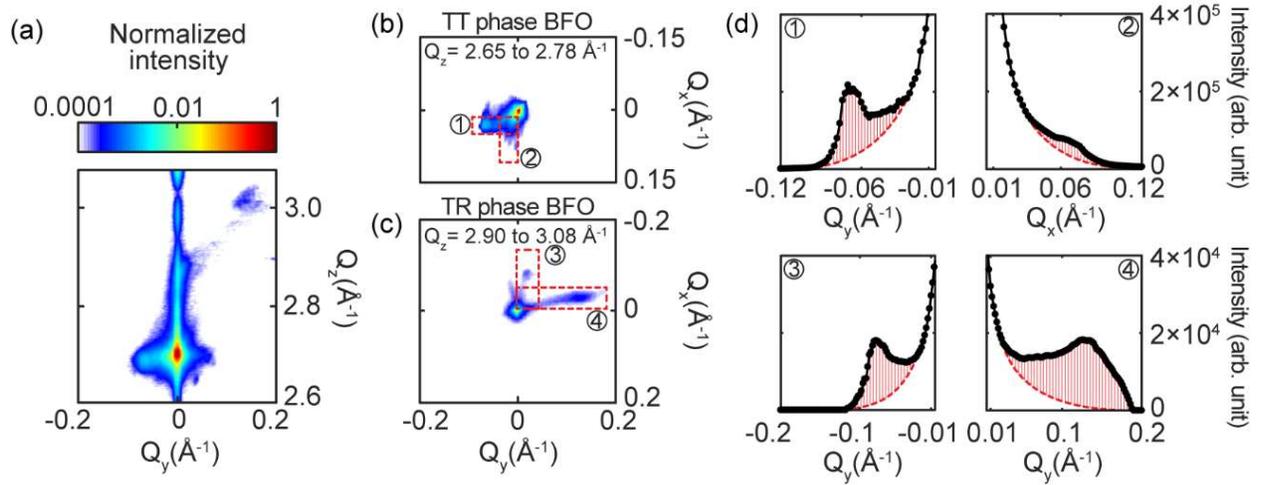

FIG S1. (a) Section of reciprocal space near the 002 T phase reflection in the $Q_y$-$Q_z$ plane of the reciprocal space at $Q_x = 0$ Å$^{-1}$. (b) $Q_x$-$Q_y$ section of reciprocal space integrating from $Q_z$=2.65 Å$^{-1}$ to $Q_z$=3.08 Å$^{-1}$. (d) Line profiles of intensity obtained by integrating reflected intensities in red boxes indicated in (b) and (c). The shaded area indicates the isolated TT and TR phase intensities by subtracting the tail of the T phase reflection.

## 2. Spatial distribution of tilted BFO phases

The spatial distribution of the tilted structural phases was measured by mapping a 100 × 100 μm$^2$ area with the microfocused x-ray beam. Figures S2(a) and (b) show maps of the integrated intensities of the TR (5) and TR (8) reflections within this region. The regions of high intensity of the TR (5) phase reflection do not overlap the regions where the TR phase (8) reflection is intense. Figs. S2(c) and (d) show maps of the diffracted intensities of the TT (4) and TT (1) reflections, respectively. The distributions of TR (5) and (8) reflections are spatially paired with TT (1) and (4) reflections, indicating the apparent coexistence of oppositely tilted TT phase and TR phase at this length scale. This microstructure of the BFO layer is consistent with previous observations of regions of alternating stripes of a TT phase with the corresponding TR phase [4-6].

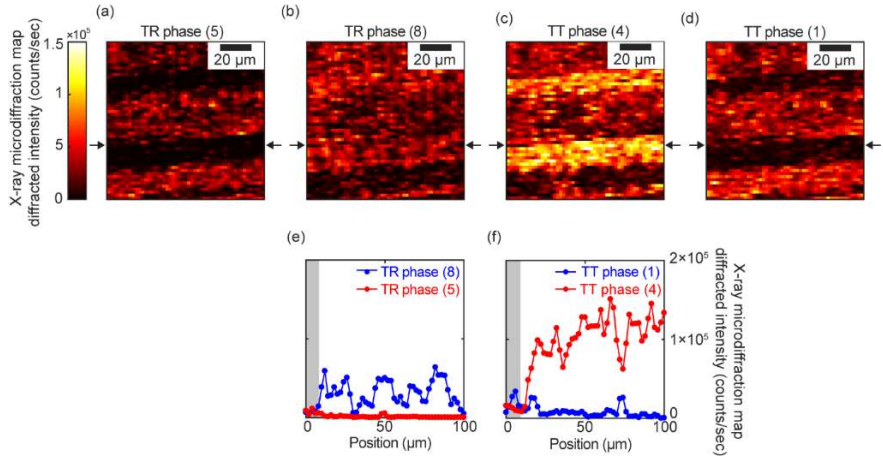

FIG S2. Maps of the integrated intensity of (a) TR (5), (b) TR (8), (c) TT (4), and (d) TT (1) phases. The probed location for the optically induced changes shown in Fig. 4 in the text is indicated by black arrows. (e) and (f) Intensity profiles of TR and TT phase reflections along the position indicated by the black arrows. The shaded area indicates the region showing a negligible optically induced change in the integrated intensity of the T phase reflection.

Intensity profiles of two TR and two TT phase reflections along the position indicated by the black arrows in the maps of Fig. S2 are shown in Figs. S2(e) and (f). The weak diffracted intensities of the tilted phase reflections in the shaded region from 0 μm to 10 μm indicates that the population of the TT and TR phases is small within that area. In addition, the T-like phase in the same region has a large lattice parameter, as in Fig. 4(b) of the text, indicating that the substrate-imposed clamping is well retained. There are negligible changes in the integrated intensity of the T phase reflection within this region following optical excitation because of the low population of the optically transformable TT and TR phases.

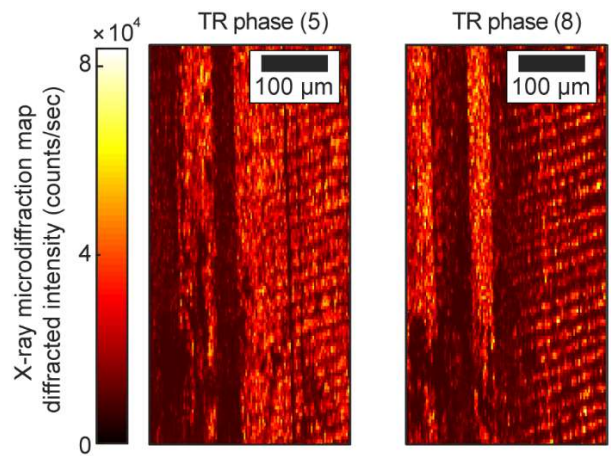

FIG S3. Maps of the integrated intensity of (a) TR (5), and (b) TR (8) phase reflections in a 300 μm × 600 μm area.

Microdiffraction maps of the integrated intensities of the TR (5) and the TR (8) reflections in a 300 μm × 600 μm area are shown in Figs. S3(a) and (b). These maps confirm that the spatial variation of the TR phases follows a similar variation throughout the entire sample. The intensity distribution of TR phase reflections explicitly shows that TR (5) phase and TR (8) phase exists in multiple spatially separated regions.

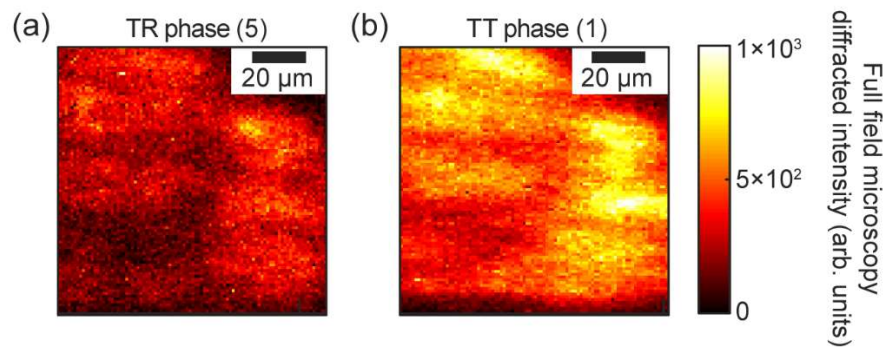

FIG S4. Full-field x-ray microscopy of (a) TR (5) and (b) TT (1) phase reflections in a 100 μm × 100 μm area.

The full-field microscopy images in Fig. S4 show that the TR (5) and TT (1) reflections have intense intensity in the same spots. The qualitative difference between the microdiffraction maps and full-field microscopy images arises because of the greater sensitivity of full-field microscopy to strain and lattice tilt under these imaging conditions.

## 3. Thermodynamic model of epitaxial thin BiFeO$_3$ film

A thermodynamic model employing the Landau-Devonshire theory was used to provide insight into the effect of the lattice expansion along the out-of-plane direction, $\Delta\varepsilon_{out}$, on the phase transformation between T and R phases. For the mechanical conditions in this study, Gibbs free energy $G$ can be expressed as [7,8]:

$$\tilde{G} = G + \sum_{n=1}^{6} u_n \sigma_n$$

Here $u_n$ and $\sigma_n$ are the strain and stress components, respectively, with indices following the Voigt notation. The standard elastic Gibbs function, $G$, is taken from Ref. [7]. Under the mechanical condition of a thin film epitaxially grown on a thick cubic substrate, the strain condition is $u_1 = u_2 = u_m$, where $u_m$ is the misfit strain arising from the difference between in-plane lattice constants of substrate and thin film. The shear stresses are zero, so that $\sigma_4 = \sigma_5 = 0$ and $u_6 = 0$. Since strain and stress along the out-of-plane direction vary in this study, $u_3 \neq 0$ and $\sigma_3 \neq 0$. Under these mechanical conditions, the modified thermodynamic potential, $\tilde{G}$, can be written as:

$$\tilde{G}(P, u_m, \sigma_3) = \alpha_1^*(P_1^2 + P_2^2) + \alpha_3^* P_3^2 + \alpha_{11}^*(P_1^4 + P_2^4) + \alpha_{33}^* P_3^4 + \alpha_{12}^* P_1^2 P_2^2 + \alpha_{13}^*(P_1^2 + P_2^2)P_3^2 + (u_m^2 - s_{12}\sigma_3^2)/(s_{11} + s_{12}) + s_{11}\sigma_3^2/2$$

Here $P_i$ is the polarization along direction $i$, $\alpha_i^*$ and $\alpha_{ij}^*$ are renormalized Landau coefficients by the misfit strain and the stress along the out-of-plane direction, $s_{ij}$ is the elastic compliance, $u_m$ is the biaxial misfit strain induced by the film-substrate lattice mismatch, and $\sigma_3$ is the applied stress along the out-of-plane direction. The renormalized Landau coefficients are:

$$\alpha_1^* = \alpha_1 + (Q_{11} + Q_{12})/(s_{11} + s_{12})u_m, \quad \alpha_3^* = \alpha_1 - 2Q_{12}[u_m - 2s_{12}^2\sigma_3/(s_{11} + s_{12})]/(s_{11} - s_{12})$$

$$\alpha_{11}^* = \alpha_{11} + [(Q_{11}^2 + Q_{12}^2)s_{11}/2 - Q_{11}Q_{12}s_{12}]/(s_{11}^2 - s_{12}^2), \quad \alpha_{33}^* = \alpha_{11} + s_{11}Q_{12}^2/(s_{11} - s_{12})^2$$

$$\alpha_{12}^* = \alpha_{12} - [(Q_{11}^2 + Q_{12}^2)s_{12} - 2Q_{11}Q_{12}s_{11}]/(s_{11}^2 - s_{12}^2) + Q_{44}^2/(2s_{44}), \text{ and } \alpha_{13}^* = \alpha_{12} + Q_{12}(Q_{11} + Q_{12})/(s_{11} - s_{12})$$

Here $\alpha_i$, and $\alpha_{ij}$ are the dielectric stiffness coefficients at constant stress, $Q_{ij}$ are the electrostrictive constants, and $s_{ij}$ are the elastic compliances at constant polarization. The list of the material parameters used in the calculations is given in Ref. [9]. The external and internal electric fields are assumed to be negligible.

The spontaneous polarization under different misfit strain and stress conditions was calculated using the condition that $\partial \tilde{G}/\partial P_i = 0$. The stress along the out-of-plane direction was varied to elongate the out-of-plane lattice parameter under the constant misfit strain condition. The lattice expansion along the out-of-plane direction was derived using $\varepsilon_3 = -\partial G/\partial \sigma_3$. The polarizations of the T and R phases are confined to the (010) plane ($|P_1| = 0, |P_2| \neq |P_3| \neq 0$) and the (110) plane ($|P_1| = |P_2| \neq 0, |P_3| \neq 0$), respectively, as described in Ref. [10].

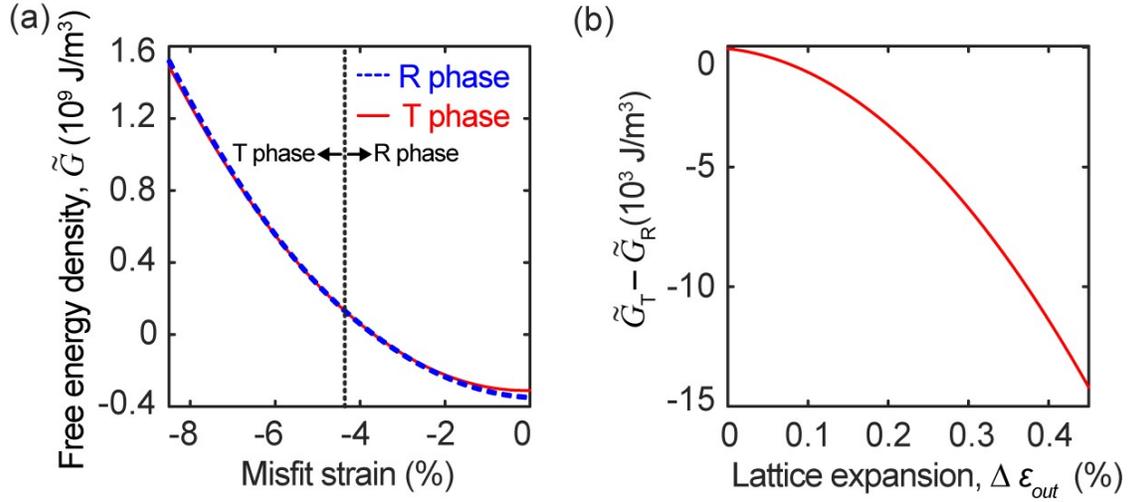

**FIG S5. (a) Free energy density, $\tilde{G}$, of R and T phases as a function of misfit strain. The black dotted line at a misfit strain of -4.3% indicates the phase boundary condition between the T and R phase. (b) Dependence of the difference between T phase free energy density, $\tilde{G}_T$, and R phase free energy density, $\tilde{G}_R$, on the out-of-plane lattice expansion at the constant misfit strain of -4.3%.**

The total free energy densities for T phase and R phase are plotted in Fig. S5(a) as a function of biaxial misfit strain under the zero out-of-plane stress condition. The energy densities of T and R phases at a misfit strain of -4.3% are equal indicating a phase boundary condition that is consistent with the reported misfit strain at which the phases are in equilibrium [4]. At a larger compressive misfit strain, T phase is stable. The R phase is energetically favorable at a strain less than -4.3%.

The difference between the free energies of the T and R phases at a misfit strain of -4.3% is plotted in Fig. S5(b) as a function of the out-of-plane strain ranging from 0 to 0.45%. A constant misfit strain was used in the calculation of Fig. S5(b) because the substrate lattice parameter is constant in the optical pump experiment [11]. The free energy density of R phase is larger than that of T phase for all values of the out-of-plane lattice expansion and, thus, the transformation from R phase to T phase is energetically favorable during expansion. The monotonic increase in the magnitude of the energy difference between the T and R phases with increasing lattice expansion in Fig. S5(b) is consistent with the experimental observation that the degree of the transformation is larger at higher out-of-plane lattice expansions. The results of this free energy model are also consistent with a previous DFT calculation [12].

**4. Optical fluence definition**

The results here are reported in terms of incident optical fluence, which we define as $F_{In}=P_{In}/(A \cdot f)$ where $F_{In}$ is the incident fluence, $P_{In}$ is the incident optical power, $A$ is the area at the FWHM of the optical power on the sample surface, and $f$ is the optical pulse repetition frequency. Other fluence definitions, for example the absorbed fluence, are inappropriate for the present experiments because the T and R phases have different absorption lengths for 400 nm wavelength light: 76 nm and 32 nm, respectively [13,14].

**5. Thermal diffusion**

A one-dimensional thermal diffusion equation was numerically solved to investigate the effects of the thermal diffusion on the dynamics of the lattice expansion. The simulation method is described in Ref. [15]. The interface heat transfer coefficient of the BFO/LAO interface is not available in the literature and we have thus used the value of 0.8 GW/(m$^2$K) reported by Wilson *et al*. for the SrRuO$_3$/SrTiO$_3$ interface [16]. The thermal diffusivity of BFO and LAO are taken from values given in Ref. [11] and Ref. [17], respectively. The optical absorption length was selected to match the R phase, in which the absorption length is less than the T phase and where, thus, effects due to thermal inhomogeneity can be expected to be larger than in the T phase.

The simulated temperature of the BFO film is plotted for several times in Fig. S6(a). The average lattice parameter, as reported in the manuscript in Fig. 2(a), depends on the depth-averaged temperature of the film, which is shown as a function of time in Fig. S6(b). Figure S6(b) shows that the heat diffuses sufficiently fast that the peak average temperature is reached within 100 ps after excitation. The slow rise time component of the expansion is thus not result from thermal diffusion within the BFO film.

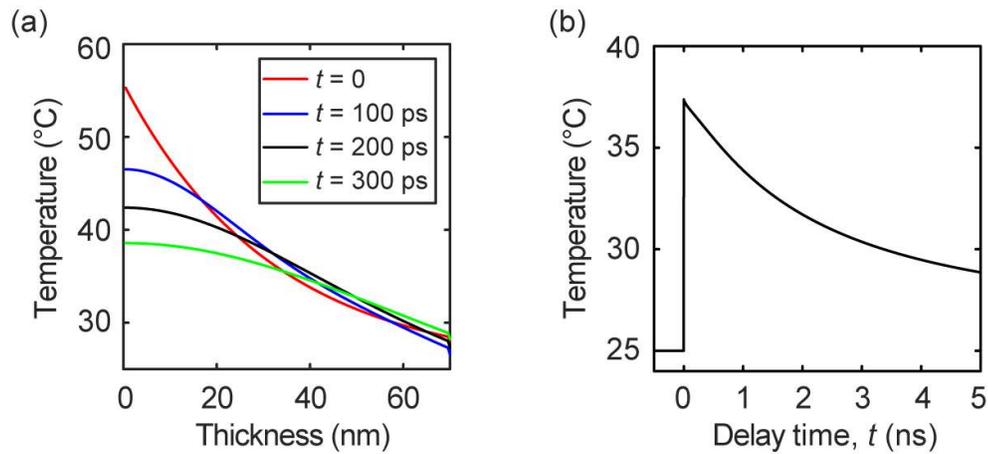

FIG S6. (a) Simulated temperature profiles within the epitaxial BFO/LAO system following optical excitation at $t=0$. (b) Time dependence of the simulated average temperature of the BFO thin film.

### 6. Normalized integrated intensities of T, TR, and TT phases

Figure S7 shows the normalized integrated intensities of the T, TR, and TT phases as a function of time following optical excitation at $t=0$. Figure S7 was assembled by plotting the intensities shown in Fig. 2b after a normalization process in which the maximum value for each reflection is set to 1. The intensities of all of the phases have the same time dependence, consistent with the variation expected in the proposed transformation model described in the manuscript.

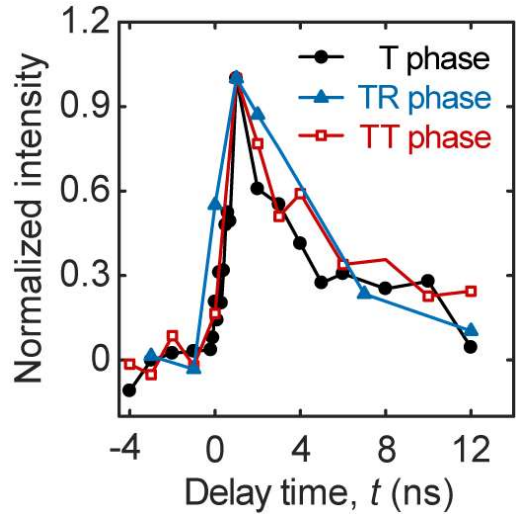

FIG. S7: Time dependence of the normalized integrated intensities of the T, TR, and TT phases.